**Dissecting Multifractal detrended cross-correlation analysis**

Borko Stosic and Tatijana Stosic

Departamento de Estatística e Informática, Universidade Federal Rural de Pernambuco, Rua Manoel de Medeiros, s/n - Dois Irmãos, 52171-900, Recife, Brazil

**Abstract**

In this work we address the question of the Multifractal detrended cross-correlation analysis method that has been subject to some controversies since its inception almost two decades ago. To this end we propose several new options to deal with negative cross-covariance among two time series, that may serve to construct a more robust view of the multifractal spectrum among them. We compare these novel options with the proposals already existing in the literature, and we provide fast code in C, with wrapper code for R and Python, for both new and the already existing proposals. We test different algorithms on synthetic series with an exact analytical solution, uncorrelated white noise series, and on daily price series of ethanol and sugar in Brazil from 2010 to 2023.

## 1 Introduction

Multifractal detrended fluctuation analysis (MFDFA) was introduced [1] two decades ago to address the multifractal behavior of nonstationary time series and has since been used in many works in a wide spectrum of areas of knowledge. The Multifractal detrended cross-correlation analysis, dubbed MF-DXA, was introduced several years later [2] to account for the power-law cross-correlations between two time series, and has since been employed in a number of works (see e.g. [3-5]). However, while the method works well when the two series are strongly correlated, it often presents serious problems when the residues of the two series (difference from the fitted polynomial trend) assume different signs. To deal with this problem, most of the works (see e.g. [6-8]) use the absolute value of the product of residues in segments of different sizes, disregarding the sign of cross-correlations, and thus violating the concept of cross-correlations in view of the original proposal [2].

To remedy this situation, a decade ago the MFCCA method was introduced [9], where the fluctuation sign is extracted before scaling, and then applied back after summation over the segments. While this proposal recovers the concept of correlations and guarantees that no imaginary terms can be encountered, in practice it can still suffer some problems due to canceling out of positive and negative terms in the sums for the segment detrended covariance and the qth order fluctuation function.

To address this issue, here we propose a set of algorithms that deal with the residue sign issue, providing grounds for a more profound assessment of power-law cross-correlations between two time series. We substantiate these proposals through a comprehensive assessment of a synthetic dataset used in [11], uncorrelated white noise series, and daily sugar and ethanol price data in Brazil from 2010 to 2023.

## 2 Methodology

### *2.1 Multifractal detrended fluctuation analysis*

Multifractal detrended fluctuation analysis (MFDFA) was introduced by Kantelhardt et al. [1] to analyze autocorrelations in nonstationary temporal series. The MFDFA algorithm starts with the integration of the original time series $x(i), i = 1, \ldots, N$, to produce the profile $X(k) = \sum_{i=1}^{k}[x(i) - \langle x \rangle]$ , $k = 1, \ldots, N$ , where $\langle x \rangle = 1/N \sum_{i=1}^{N} x(i)$ is the mean. Next, the series $X(k)$ is divided into $N_n$ segments of length $n$, and in each segment $\nu = 1, \ldots, N_n$ the local trend $X_{n,\nu}$ is obtained through a linear or higher order polynomial least square fit. The so-called detrended variance

$$F^2(n,\nu) = \frac{1}{n} \sum_{k=(\nu-1)n+1}^{\nu n} \left(X(k) - X_{n,\nu}(k)\right)^2 \tag{1}$$

is calculated for each segment and used to find the $q$th order fluctuation function

$$F_q(n) = \left\{ \frac{1}{N_n} \sum_{\nu=1}^{N_n} \left(F^2(n,\nu)\right)^{q/2} \right\}^{1/q} \tag{2}$$

where $q$ is a parameter that can take on any real value except zero.

Repeating this procedure for different segment sizes provides the relationship between the fluctuation function $F_q(n)$ and the box size $n$. If long-term correlations are present, the fluctuation function increases with $n$ as a power law $F_q(n) \sim n^{h(q)}$, and if there are no correlations $F_q(n) \sim n^{1/2}$ for all $q$ values, characteristic of random walk. The scaling exponent $h(q)$ is called the generalized Hurst exponent and is obtained as the slope of linear regression of $\log F_q(n)$ versus $\log n$. For stationary time series $h(2)$ corresponds to the classical Hurst exponent $H$. The exponents $h(q)$ describe the scaling behavior of subsets of series with large fluctuations (for positive $q$ values) and subsets with small fluctuations (for negative $q$ values). If subsets with small and large fluctuations scale differently and $h(q)$ is a decreasing function of $q$ the underlying process is multifractal, while for monofractal time series, $h(q)$ is constant. The range of $h(q)$ values can be used to describe the degree of multifractality, so that the series that generates larger range of $h(q)$ is considered to display stronger multifractality.

It is often more convenient to describe the multifractality of a time series by using the properties of the multifractal spectrum $f(\alpha)$ which is obtained through the Legendre transform

$$\alpha(q) = \frac{d\tau(q)}{dq}, \tag{3}$$

$$f(\alpha(q)) = q\alpha(q) - \tau(q), \tag{4}$$

where $\tau(q) = qh(q) - 1$. Monofractal process is here represented by a single point in the $f(\alpha)$ plane, while a multifractal process generates a single humped function [1].

Multifractal spectrum $f(\alpha)$ contains the information about complexity of the underlying stochastic process, trough three complexity parameters: the position of the maximum $\alpha_0$, the width of the spectrum $\Delta\alpha = \alpha_{max} - \alpha_{min}$, and the skew parameter $r = (\alpha_{max} + \alpha_{min} - 2\alpha_0)/(\alpha_{max} - \alpha_{min})$ where $-1 \leq r \leq 1$ ($r = 0$ for symmetric shapes, $r > 0$ for right-skewed shapes, and $r < 0$ for left-skewed shapes) [12,13]. If $\alpha_0 > 0.5$ the underlying process is overall persistent (larger value of $\alpha_0$ indicates stronger persistency), and if $\alpha_0 < 0.5$ the process is overall antipersistent (smaller value of $\alpha_0$ indicates stronger antipersistency). The width $\Delta\alpha$ of the spectrum measures the degree of multifractality of the process (larger $\Delta\alpha$ value indicates stronger multifractality). The skew parameter $r$ indicates the dominance of small ($r > 0$) or large ($r < 0$) fluctuations in multifractality of the process [12,13].

*2.2 Multifractal detrended cross-correlation analysis*

Multifractal detrended cross-correlation fluctuation analysis (MF-DXA) was introduced by Zhou [2] as a generalization of MFDFA for cross-correlation of two time series $x(i)$ and $y(i), i = 1, \dots, N$. Dividing the profiles $X(k)$ and $Y(k)$ into $N_n$ segments of length $n$, and finding the local trends $X_{n,\nu}$ and $Y_{n,\nu}$ for segments $\nu = 1, \dots, N_n$ through least square fitting the detrended covariance is calculated as

$$F_{xy}^2(n,\nu) = \frac{1}{n} \sum_{k=(\nu-1)n+1}^{\nu n} \left(X(k) - X_{n,\nu}(k)\right)\left(Y(k) - Y_{n,\nu}(k)\right) \qquad (5)$$

and the $q$th order fluctuation function as

$$F_q(n) = \left\{ \frac{1}{N_n} \sum_{\nu=1}^{N_n} \left(F_{xy}^2(n,\nu)\right)^{q/2} \right\}^{1/q} \quad . \qquad (6)$$

The problem with this approach is that $F_{xy}^2(n,\nu)$ can be negative in which case $F_q(n)$ is not well defined. To deal with this problem majority of works [6-8] implement absolute values of the residue products in (5) as

$$F_{ABS,xy}^2(n,\nu) = \frac{1}{n} \sum_{k=(\nu-1)n+1}^{\nu n} \left| \left(X(k) - X_{n,\nu}(k)\right)\left(Y(k) - Y_{n,\nu}(k)\right) \right| \quad , \qquad (7)$$

but this approach (henceforth referred to as ABS) may distort or even spuriously amplify the multifractal cross-correlation measures [9].

Another approach named MFCCA was proposed [9,10] to account for the sign of the cross-correlation through

$$F_{xy}^q(n) = \frac{1}{N_n} \sum_{\nu=1}^{N_n} sgn\left(F_{xy}^2(n,\nu)\right) \left|F_{xy}^2(n,\nu)\right|^{q/2} , \qquad (8)$$

and

$$F_{qxy}(n) = sgn\left(F_{xy}^q(n)\right) \left|F_{xy}^q(n)\right|^{1/q} \quad . \qquad (9)$$

This approach guarantees that no imaginary terms can be encountered, and it has been implemented in a number of works [14-16]. Nevertheless, the positive and negative values in (5) and (8) can cancel out, which may pose a problem in some situations.

***2.3 The new proposal***

In what follows we present several ways to deal with negative cross correlations among the series. The first approach is to separate the positive and the negative values of the detrended covariance $F_{xy}^2(n,\nu^+) > 0$ and $F_{xy}^2(n,\nu^-) < 0$. If there are $N_n^+$ and $N_n^-$ segments of size $n$ with positive and negative detrended covariance, respectively, such that $N_n^+ + N_n^- = N_n$, two fluctuation functions are obtained as

$$F_q^+(n) = \left\{\frac{1}{N_n^+}\sum_{\nu^+=1}^{N_n^+}\left(F_{xy}^2(n,\nu^+)\right)^{q/2}\right\}^{1/q} \quad , \quad F_q^-(n) = \left\{\frac{1}{N_n^-}\sum_{\nu^-=1}^{N_n^-}\left(-F_{xy}^2(n,\nu^-)\right)^{q/2}\right\}^{1/q} \quad , \qquad (9)$$

resulting in two multifractal spectra. If the two series are strongly correlated (or strongly anticorrelated) the number of segments with negative (positive) detrended covariance is small, and the resulting spectrum should be discarded. This approach shall be referred to henceforth as Plus sum (PS) and Minus sum (MS).

The next approach is to separate the $n^+$ positive from the $n^-$ negative residue products already inside the sum of equation (5) over the segment. Using notation $\varepsilon_{xk} = X(k) - X_{n,\nu}(k)$ and $\varepsilon_{yk} = Y(k) - Y_{n,\nu}(k)$ the two detrended covariance expressions are now given by

$$F_{xy}^2(n^+,\nu) = \frac{1}{n^+}\sum_{\substack{k=(\nu-1)n+1 \\ \varepsilon_{xk}\varepsilon_{yk}>0}}^{\nu n} \varepsilon_{xk}\varepsilon_{yk} \quad , \quad F_{xy}^2(n^-,\nu) = -\frac{1}{n^-}\sum_{\substack{k=(\nu-1)n+1 \\ \varepsilon_{xk}\varepsilon_{yk}<0}}^{\nu n} \varepsilon_{xk}\varepsilon_{yk} \quad , \qquad (10)$$

again producing two multifractal spectra. This approach shall be referred to henceforth as Plus box (PB) and Minus box (MB).

Finally, breaking down further the contributions to the detrended covariance of point pairs with $\varepsilon_{xk} > 0$, $\varepsilon_{yk} > 0$ (PP version), $\varepsilon_{xk} > 0$, $\varepsilon_{yk} < 0$ (PM version), $\varepsilon_{xk} < 0$, $\varepsilon_{yk} > 0$ (MP version), and $\varepsilon_{xk} < 0$, $\varepsilon_{yk} < 0$ (MM version), yields four multifractal spectra, which may turn out helpful in discerning the origin of cross-correlation multifractality.

## 3 Test results

In what follows we test different algorithms of the multifractal cross correlation analysis on the Binomial multifractal model which has an exact theoretical solution, then on uncorrelated white noise series, and on data for daily sugar and ethanol prices in Brazil for the period 2010 to 2023. To this end we use the code written in C, with wrappers for R and Python, available at https://github.com/borkostosic/MFDCCALIB.

### *3.1 Binomial multifractal model*

Binomial multifractal model is well studied, with an analytical solution [11], thus providing solid grounds for testing numerical algorithms. It was also used in [1] where the MFDFA method was proposed, and in [2] where the MF-DXA method was introduced.

The model assigns a measure to $2^n$ segments of the unit interval through an n-stage recursive multiplicative process. At stage $n = 1$ the unit interval is divided into two parts; the left part is attributed measure $0 < p < 1$ and the right part measure $(1 - p)$. At each of the following stages $k = 2, \dots, n$ each of the $2^k$ segments is divided in two, and the left and the right parts are attributed a product of the measure of the original segments at stage $k - 1$ multiplied by $p$ and $(1 - p)$, respectively (see [1,2] for more details). The resulting set of numbers corresponding to the measure of the $2^n$ segments are calculated as $x_i = p^{b(i-1)}(1 - p)^{n-b(i-1)}$ where $b(\cdot)$ is the number of 1's in the binary representation of the argument. The multifractal measures in the limit $n \to \infty$ are given by expressions

$$H(q) = \frac{1}{q}\{1 - \log_2[p^q + (1-p)^q]\} \ ,$$

$$\tau(q) = -\log_2[p^q + (1-p)^q] \ ,$$

$$\alpha(q) = \frac{p^q \log_2 p + (1-p)^q \log_2(1-p)}{p^q + (1-p)^q} \ ,$$

$$f(q) = q\frac{p^q \log_2 p + (1-p)^q \log_2(1-p)}{p^q + (1-p)^q} + \log_2[p^q + (1-p)^q] \ . \qquad (11)$$

In Fig.1 the results of the MFDFA and MF-DXA algorithms for $p = 0.3$ and $p = 0.4$ (the same choice as in [2]) are shown for $n = 17$ (two sequences of $2^{17} = 131072$ numbers each). It is seen from Fig. 1 that the exact theoretical values are well reproduced for $p = 0.3$, somewhat less so for $p = 0.4$, and the average between $p = 0.3$ and $p = 0.4$ rather well corresponds to the MF-DXA numerical results.

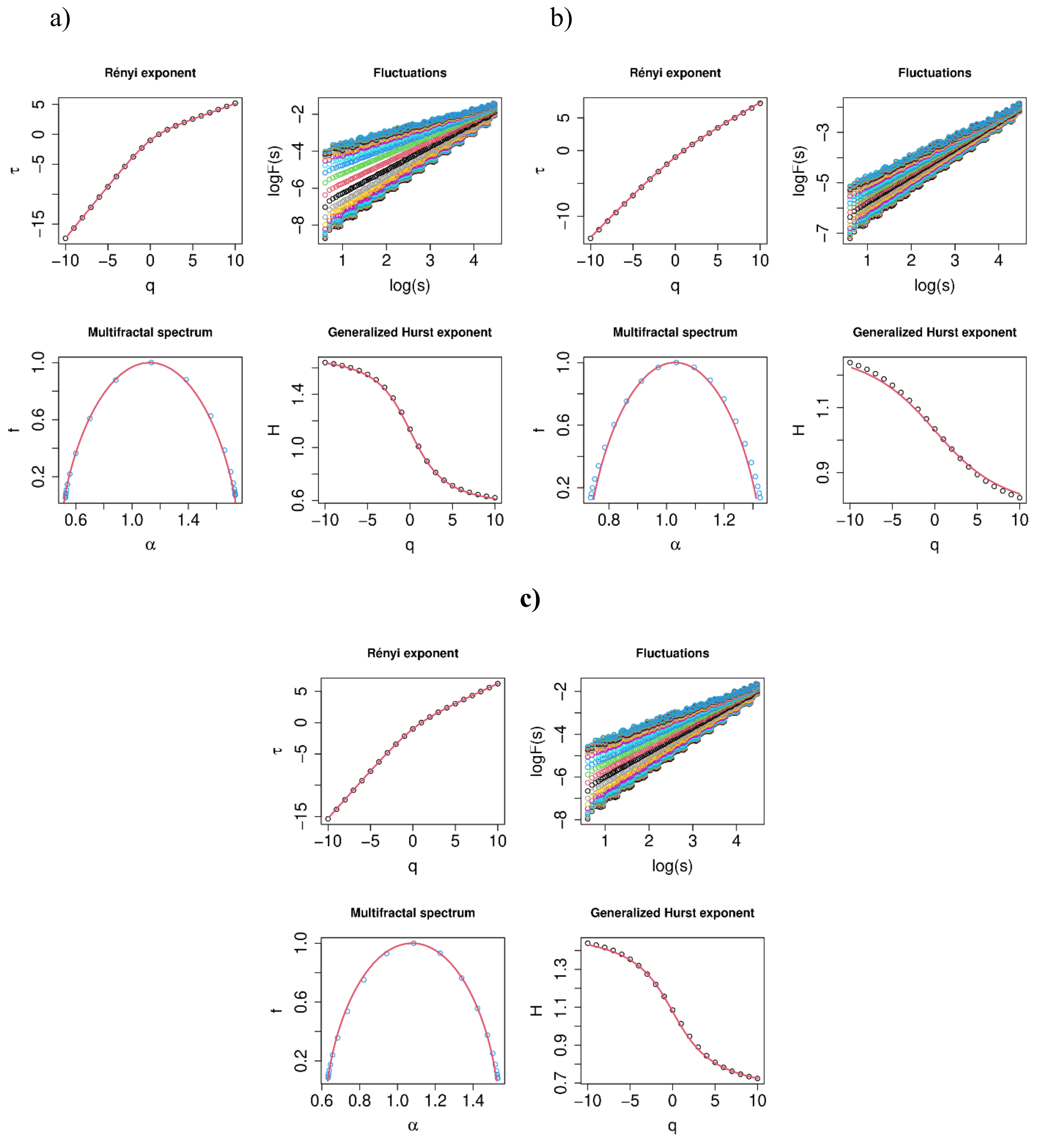

**Figure 1.** MFDFA simulation results together with theoretical curves for a) $p = 0.3$ and b) $p = 0.4$, together with c) the MF-DXA numerical results. The red curves represent theoretical results in a) and b), and in c) they correspond to the average of the theoretical curves in a) and b).

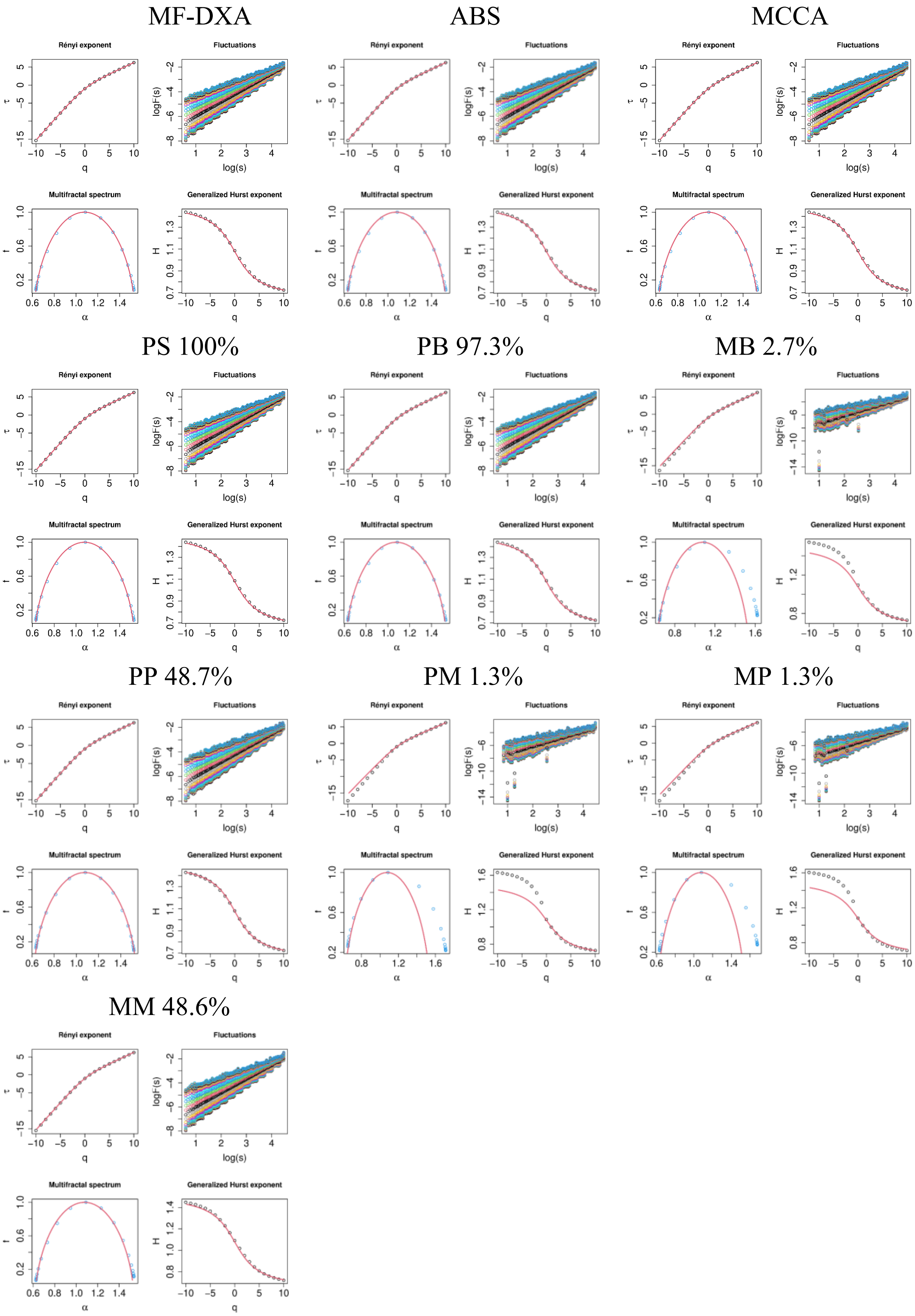

**Figure 2.** MFDCCA simulation results for the Binomial multifractal model for $p = 0.3$ and $p = 0.4$ using different algorithms (see text for details). The numbers next to the labels represent the percentage of point pairs satisfying the criterion of the corresponding algorithm, and the red curves correspond to the average of the theoretical MFDFA curves for the two series.

The results of the different algorithms from the previous Section, MF-DXA, ABS, MFCCA, PS, PB, MB, PP, PM, MP and MM are presented in Fig. 2 and Tab. 1, for the Binomial multifractal model with $p = 0.3$ and $p = 0.4$, for $n = 17$ (sequences of $2^{17} = 131072$ numbers). The MS is not shown in Fig. 2 as no segments with negative detrended covariance $F_{xy}^2(n, \nu^-)$ were found, for any segment size. The numbers next to the labels for PS, PB, MB, PP, PM, MP and MM in Fig. 2 represent the overall percentage of point pairs satisfying the criterion of the corresponding algorithm, also presented in the second column of Tab. 1. It is seen in Fig. 2 that MF-DXA, ABS, MFCCA, PS, PB, PP and MM yield practically identical results, very close to the average of the theoretical MFDFA curves for the two series, even if PB, PP and MM were satisfied for 90.8%, 38.7% and 52.1% of point pairs, respectively.

**Table 1.** Multifractal spectra parameters (Hurst exponent $H$, the position of the maximum $\alpha_0$, spectrum width $\Delta\alpha$, and the skew parameter $r$) for the Binomial multifractal model for $p = 0.3$ and $p = 0.4$ using different algorithms.

| Algorithm | Pairs (%) | $H$ | $\alpha_0$ | $\Delta\alpha$ | r |
|---|---|---|---|---|---|
| Theoretical | - | 0.932 | 1.078 | 0.904 | 0 |
| MFDFA average | - | 0.934 | 1.086 | 0.8974 | -0.0123 |
| MFDXA | 100 | 0.946 | 1.086 | 0.8981 | -0.009 |
| ABS | 100 | 0.947 | 1.086 | 0.8981 | -0.009 |
| MFCCA | 100 | 0.946 | 1.086 | 0.8981 | -0.009 |
| PS | 100 | 0.946 | 1.086 | 0.8981 | -0.009 |
| MS | 0 | | | | |
| PB | 97.3 | 0.947 | 1.086 | 0.8992 | -0.009 |
| MB | 2.7 | 0.949 | 1.096 | 0.9734 | 0.077 |
| PP | 48.7 | 0.940 | 1.081 | 0.8841 | -0.008 |
| PM | 1.3 | 0.927 | 1.086 | 1.0587 | 0.181 |
| MP | 1.3 | 0.927 | 1.080 | 1.0447 | 0.141 |
| MM | 48.6 | 0.951 | 1.091 | 0.9108 | -0.020 |

The algorithms where only a small fraction of point pairs satisfy the corresponding criteria (MB with 2.7%, PM with 1.3%, and MP with 1.3%) present some serious distortions of the fluctuations at smaller segment sizes, resulting in wrong multifractal spectra parameters. While this may be remedied by using a smaller range of segment sizes to avoid smaller segments (for the results in Fig.2 segment sizes of 4 to $2^{17}/4 = 32768$ were used), it follows that results of an algorithm with a low percentage of point pairs satisfying the corresponding criterion should be avoided. It seems prudent to run all the algorithms to assess the different aspects of cross-correlations of the two series at different scales, before choosing the most consistent options.

### *3.2 White noise*

We have applied all the algorithms (the ones in the existing literature, and the ones proposed in this work) on two independent white noise series of size $2^{17}$=131072 with the results presented in Fig. 3. Runing the MFDFA algorithm on the two series independently produces generalized Hurst exponent $H(q)\sim 0.5$ and the multifractal spectrum at a single point $f(0.5) = 1$, as expected for random walk. As covariance between the two series has an equal chance of assuming positive and negative values the MF-DXA algorithm produces no results at all, and the MFCCA algorithm yields (for roughly half of the segments where the q-th order fluctuation function happens to be positive) the $H(q)$ function that diverges close to $q\sim 0$, and a cloud of points for the $f(\alpha)$ spectrum.

While there are no cross-correlations among the two series, there are no correlations in the original individual series either, and one may expect that the absolute values of cross-correlations and the subsets of only positive and only negative cross-correlations should also follow random walk behavior. From Fig. 3 it follows that such behavior is captured by the ABS, PB and MB algorithms, while PS and MS suffer from canceling out of terms in summation of equation (5) resulting in a spurious right-hand side of the spectrum (spurious small fluctuations). The same effect is also observed to a lesser extent for PP, PM, MP and MM algorithms.

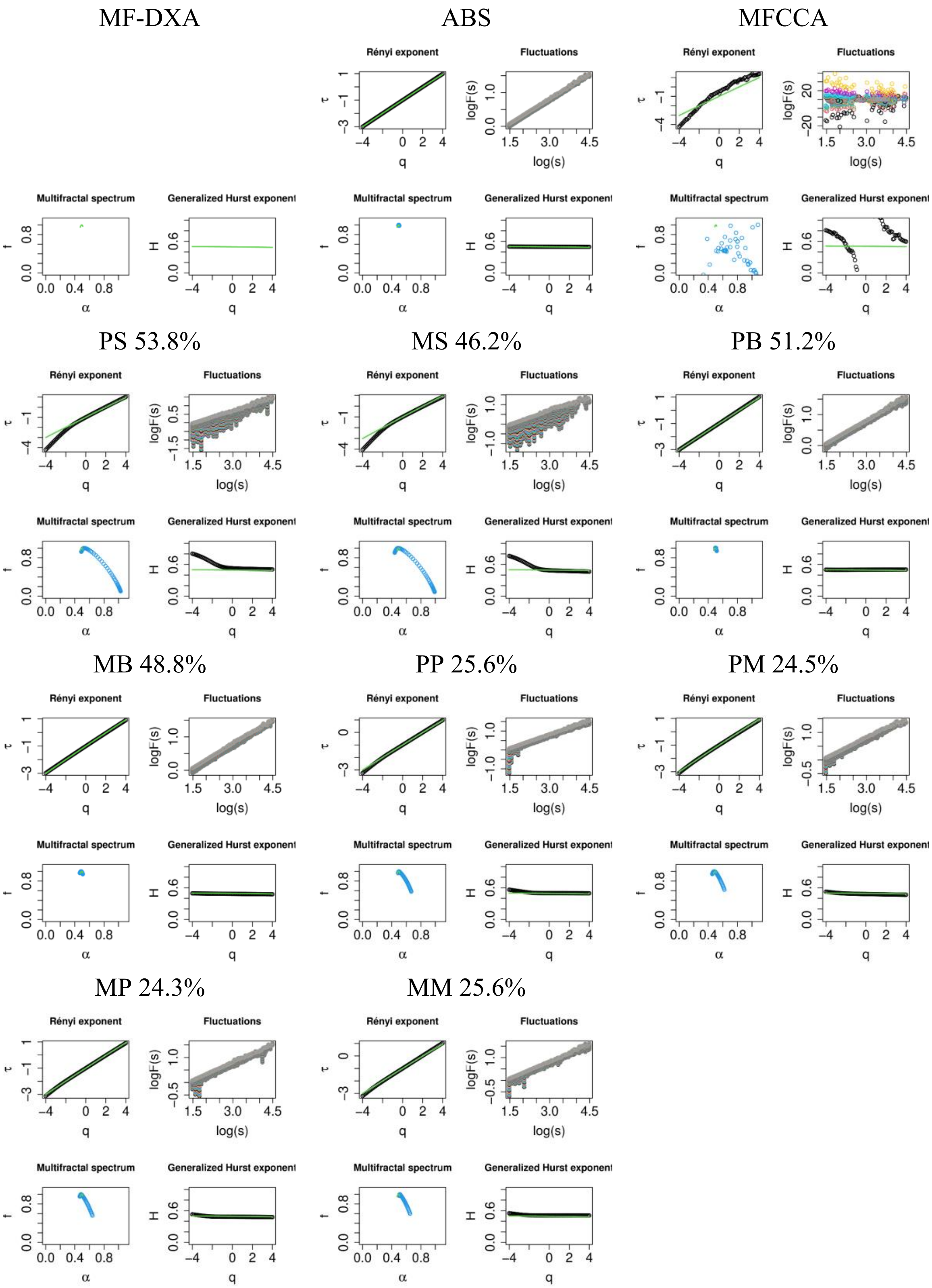

**Figure 3.** MFDCCA simulation results for two uncorrelated white noise series. The numbers next to the labels represent the percentage of point pairs satisfying the criterion of the corresponding algorithm, and the green curves correspond to the average of the MFDFA curves for the two series.

### *3.3 Sugar ethanol prices in Brazil from 2010 to 2023*

Sugar and ethanol are produced in Brazil from sugar cane, where technological installations make it possible to quickly switch between production of these commodities [17], depending on the demand and price. The historical sequences of these prices therefore represent an interesting platform for empirical studies, important from the economic point of view. The data were obtained from the CEPEA/ESALQ site https://www.cepea.esalq.usp.br/br (CEPEA - Centro de Estudos Avançados em Economia Aplicada, Departamento de Economia, Administração e Sociologia, ESALQ - Escola Superior de Agricultura Luiz de Queiroz, USP - Universidade de São Paulo). In what follows we address the Multifractal detrended cross-correlation analysis of the logarithmic returns of these two daily price series between 2010 and 2023, using all the above methods (those already existing in the literature, and the ones proposed in this work). The results of these calculations are shown in Fig. 4 and Tab. 2.

It is evident that MF-DXA and MFCCA yield erroneous results, the former because positive cross correlations are found only for very large segments, and the latter because of divergent behavior for low values of $q$ caused by canceling out of terms in the sums of equations (5) and (8), while the ABS version of the three existing methods yields rather "reasonable" graphs. On the other hand, all the algorithms proposed in this work yield wider, more asymmetric spectra in comparison to the ABS (modulus MF_DXA) approach. The percentage of point pairs captured by different flavors of the algorithm varies between 20.4% (PM) and 67.5% (PS), and the resulting multifractal parameters are rather consistent across the set of options, as can be seen in Tab. 2. The individual algorithm implementation could be further improved by inspecting the $\log F_q(s)$ versus $\log s$ graphs in Fig. 4 and restricting the segment size range in each case. For example, the MS algorithm captures 32.5% of point pairs, but the scaling is distorted for very large segment sizes, pushing the $h(q)$ curve downwards for positive $q$ values, and thus the $f(\alpha)$ spectrum to the left. Improved "linearity" of $\log F_q(s)$ versus $\log s$ could thus be obtained by reducing the upper bound for segment size $s$ values. The most consistent of more detailed restriction algorithms (the most "linear" of the log-log plots for different $q$ values) appears to be PP with 28.4% point pairs captured.

MF-DXA ABS MFCCA

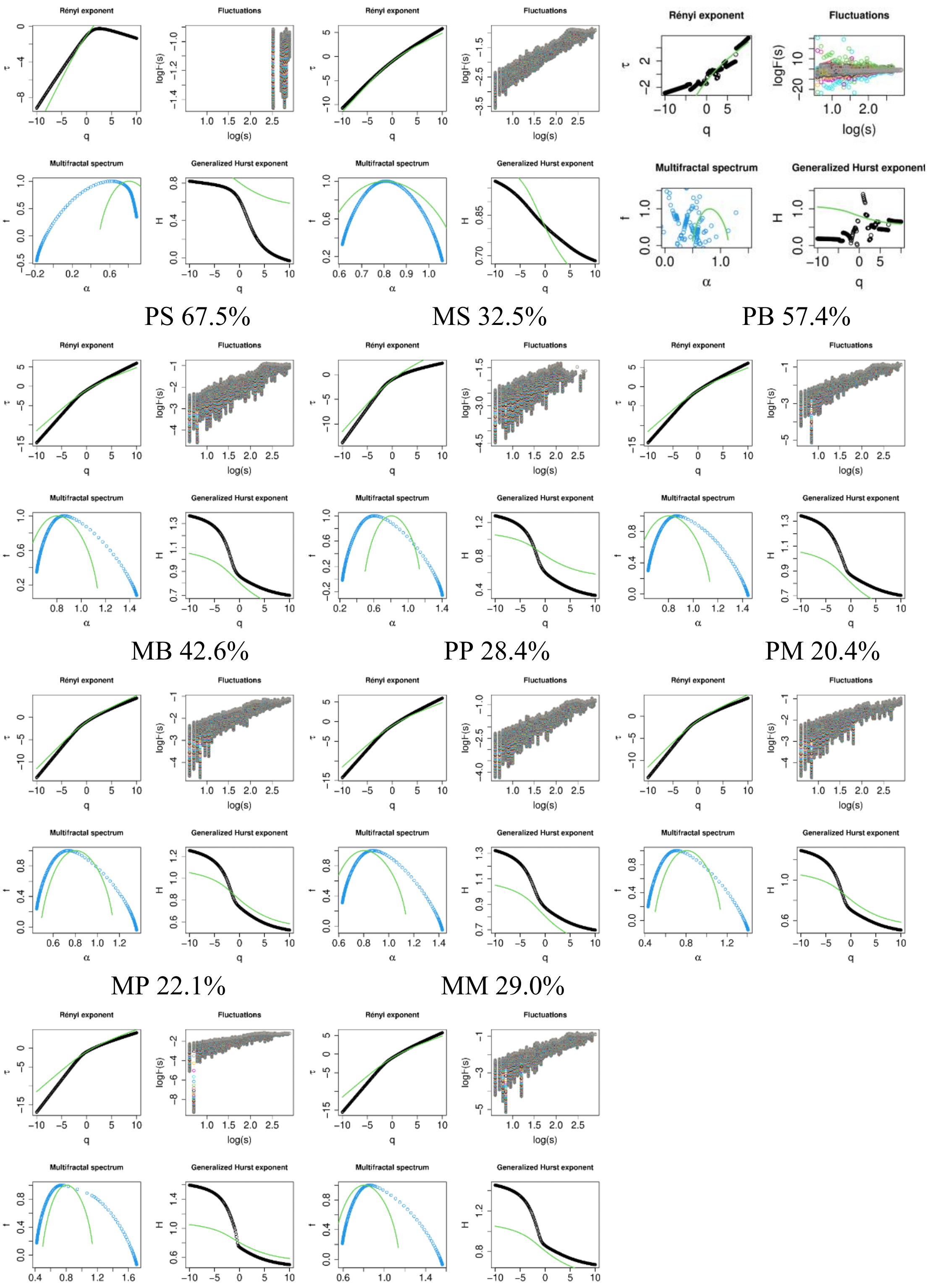

**Figure 4.** MFDCCA results for the sugar-ethanol data using the existing and the novel algorithms proposed here (see text for results). The green curves correspond to the average of the corresponding MFDFA curves for the two series.

**Table 2.** Multifractal spectra parameters for sugar-ethanol data using different algorithms.

| Algorithm | Pairs (%) | *H* | a0 | W | r |
|---|---|---|---|---|---|
| MFDFA average | - | 0.672 | 0.805 | 0.636 | 0.034 |
| MFDXA | 100 | 0.105 | 0.618 | 1.067 | -0.496 |
| ABS | 100 | 0.745 | 0.809 | 0.447 | 0.128 |
| MFCCA | 100 | 0.793 | - | - | - |
| PS | 67.5 | 0.762 | 0.861 | 0.823 | 0.449 |
| MS | 32.5 | 0.435 | 0.605 | 1.103 | 0.319 |
| PB | 57.4 | 0.773 | 0.866 | 0.808 | 0.424 |
| MB | 42.6 | 0.605 | 0.738 | 0.898 | 0.364 |
| PP | 28.4 | 0.766 | 0.875 | 0.791 | 0.378 |
| PM | 20.4 | 0.590 | 0.707 | 0.940 | 0.407 |
| MP | 22.1 | 0.585 | 0.742 | 1.247 | 0.482 |
| MM | 29.0 | 0.750 | 0.859 | 0.953 | 0.440 |

At this point one may opt for one of two approaches: i) choose the "most linear" of the algorithms (in this case PP), or ii) calculate the average of all the "reasonably linear" algorithm results, perhaps weighted by the percentage of captured point pairs. In any case, as the MFDFA as well as all the presented MFDCCA algorithms should be seen as robust scaling approximations to the behavior of the underlying process, both approaches may be seen as reasonable, as long as one avoids the miscalculation pitfalls as seen for MF-DXA and MFCCA in the current example.

## 4 Summary

In this work we propose a spectrum of algorithms to deal with the negative cross-covariance in the multifractal cross correlation analysis of two temporal series. The original method MF-DXA ignores this issue and works only if negative cross-covariance is not observed. The modulus version of MF-DXA (with an acronym ABS in the present work) resolves this issue by using the absolute value of the cross-covariance but may distort or even spuriously amplify the multifractal cross-correlation measures [9]. Finally, the MFCCA method [9,10] deals with this issue to account for the sign of the cross-correlations by extracting the sign of segment cross-covariance and applying it back after $q$-scaling, but canceling out of the positive and negative terms in equations (5) and (8) renders this method unsuitable for two uncorrelated white noise series, and the ethanol and sugar daily price series.

The novel algorithms proposed in this work are all based on the idea of separating positive from negative cross-covariances and performing $q$-scaling independently, in each group. The PS and MS algorithms perform the summation within the segments and then separate the results into the positive and the negative groups. The PB and MB algorithms are more detailed, they distinguish the positive and negative cross-covariance contributions already at the segment level, counting the number of occurrences for each sign. Finally, the PP, PM, MP, and MM algorithms go a step further by distinguishing cases when both series' pair points are found to be above the trend (PP), when the first series point lies above the trend and the second series point is below the trend (PM), when the first series point lies below the trend and the second series point is above the trend (MP), and when both series' points lie below the corresponding trend (MM). Any of these algorithms may encounter only a small percentage of point pairs satisfying the corresponding condition, in which case it should be discarded (or analyzed with more care). The conclusion is that one may: i) choose the "most linear" of the algorithms, or ii) calculate the average of all the "reasonably linear" algorithm results. To facilitate experimenting with all the above algorithms we provide source code in C, R and Python at the link https://github.com/borkostosic/MFDCCALIB.

**Acknowledgments**

The authors acknowledge support of Brazilian agency CNPq through the research grants Nº 308782/2022-4 and Nº309499/2022-4**.** B.S. acknowledges support of Brazilian agency CAPES through grant No 88887.937789/2024-00.